\documentstyle[twocolumn,aps]{revtex}
\begin{document}

\draft
\title{Plaquette Resonating-Valence-Bond Ground State of CaV$_4$O$_9$}
\author{Kazuo Ueda and Hiroshi Kontani}
\address{
Institute for Solid State Physics, University of Tokyo, \\
7-22-1 Roppongi, Minato-ku, Tokyo 106, Japan}
\author{Manfred Sigrist and Patrick A. Lee}
\address{Department of Physics, Massachussetts Institute of
Technology, \\ Cambridge, Massachussetts 02139}

\date{\today}
\maketitle
\begin{abstract}
A theoretical model is presented to explain the spin gap
observed for CaV$_4$O$_9$.  The underlying lattice of the 1/5-depleted
square lattice favors a formation of plaquette resonating
valence bond state. Inclusion of the frustrating second neighbor
interaction enhances this tendency, leading to a quantum disordered state
of a two dimensional spin-1/2 Heisenberg model with a sufficiently
big spin gap compatible with experiments.
\end{abstract}


Quantum disordered phases with a spin gap are of great current
interest.  This topic has gained additional momentum by Anderson's
proposal of the resonating valence bond (RVB) state in the undoped
parent materials of high temperture superconductivity\cite{RVB}.
Some of the typical examples with spin gaps
are the spin-1 antiferromagnetic
Heisenberg chain\cite{haldane}, the double chain spin-1/2 Heisenberg
model\cite{strong}, the spin-1/2 Heisenberg antiferromagnet
on a Kagom\'{e} lattice\cite{zeng}, and Kondo spin liquid phase of the
Kondo lattice model at half-filling\cite{tsunetsugu}.

Recently a new system with a spin gap was found experimentally
by Taniguchi et al. for CaV$_4$O$_9$ \cite{taniguchi}.  The spin gap
observed by magnetic susceptibility and nuclear
magnetic resonance (NMR) measurements is $\Delta/k_{\rm B}=107$K.
In this paper we propose
that the underlying lattice of the 1/5-depleted square lattice of
CaV$_4$O$_9$,
see Fig.1, favors a new type of spin disordered phase which may be
called as plaquette resonating-valence-bond (PRVB) state.

Each vanadium ion occupies a crystallographycally equivalent site
and is surrounded by a pyramid of oxygens.  First let us discuss
the electronic state of this cluster (VO$_5$)$^{2-}$.  In this
configuration the vanadium ion is in V$^{4+}$ with one $d$-electron.
Since V$^{4+}$ is surrounded by a pyramid of oxygen ions, the
$d$-electron is in either $d_{xz}$  or $d_{yz}$  orbital.
This two-fold
degeneracy is lifted by a small Jahn-Teller distorsion whose existence
is reported in \cite{taniguchi} but its details are not yet clear
\cite{dxy}.
However for our discussion of the spin gap, the details are not
important because V$^{4+}$ has a magnetic moment of spin-1/2
for which single ion anisotropy is absent.

The couplings between the spins on the vanadium ions are mediated
by superexchange via the oxygen orbitals.
The nearest neighbor vanadium ions share an edge of the square of oxgens
(edge sharing), while the next nearest neighbor pairs share an oxgen
at a corner (corner sharing).
Superexchanges between the spins are possible through
the hybridization with the $p_z$ orbitals of these oxgens.
Since the number of paths for the edge sharing and the corner sharing
is two and one, respectively, we expect $J_1 {\rm (edge\ sharing)} \cong
2 J_2 {\rm (corner\ sharing)}$. Thus an appropriate model for CaV$_4$O$_9$
is the spin-1/2 Heisenberg model on the 1/5-depleted square lattice
\begin{equation}
   {\cal H}=J_1\sum_{n.n.}{\bf s}_i \cdot {\bf s}_j
           +J_2\sum_{n.n.n.}{\bf s}_i \cdot {\bf s}_j \ .
\end{equation}
The magnitude of the exchange couplings may be estimated
>from the susceptibility data\cite{taniguchi}. At high
temperatures it is reasonable to assume a Curie-Weiss
behavior and the Weiss constant is given by $k_{\rm B}\theta=
\frac{1}{3}s(s+1)(z_1J_1+z_2J_2)$, where $s$ is the spin quantum
number, $z_1$ is the number of
nearest neighbors and $z_2$ the number of next nearest neighbors.
For the 1/5-depleted square lattice $z_1=z_2=3$.
 From the intersect of the inverse susceptibility with the
temperature axis it is estimated as $\theta=220$\ K\cite{sato}. Under the
assumption of $J_1=2J_2$, we obtain $J_1/k_{\rm B}\sim200$\ K.    

To understand specific features of the 1/5-depleted square lattice
let us consider a cluster of four spins
on a plaquette.  The Hamiltonian of this system is
\begin{equation}
   {\cal H}_{\rm plaquette} =J_1({\bf s}_1+{\bf s}_3)
   \cdot ({\bf s}_2+{\bf s}_4) + J_2({\bf s}_1\cdot{\bf s}_3
   +{\bf s}_2\cdot{\bf s}_4).
\end{equation}
This Hamiltonian is readily diagonalized as is shown in Table I,
where $S_{13}$ ($S_{24}$) is the spin quantum number of
${\bf s}_1+{\bf s}_3$ (${\bf s}_2+{\bf s}_4$) and
$S$ is the total spin quantum number.

\begin{center}
Table I \ \ Eigenstates of a plaquette\\

\begin{tabular}{|l|c|c|c|c|c|c|}\hline\hline
$S_{13}$ & 0 & 1 & 0 & \multicolumn{3}{c|} 1 \\
$S_{24}$ & 0 & 0 & 1 & \multicolumn{3}{c|} 1 \\ \cline{5-7}
$S     $ & 0 & 1 & 1 & 0 & 1 & 2 \\ \hline
& & & & & & \\
$E_g   $ & $-\frac{3}{2}J_2$ & $-\frac{1}{2}J_2$ & $-\frac{1}{2}J_2$
         & $-2J_1+\frac{1}{2}J_2$ & $-J_1+\frac{1}{2}J_2$ &
           $J_1+\frac{1}{2}J_2$ \\
& & & & & & \\
\hline \hline
\end{tabular}
\end{center}

\noindent
It is seen that the ground state of the plaquette
is a singlet and the first excited state
is a spin triplet with a spin gap of $J_1$ for $J_1/2 > J_2$.  At
$J_1 = 2 J_2$ the spin neutral excitation crosses with the spin
triplet excitation.

To proceed further let us first discuss a simplified Heisenberg
model on the 1/5-depleted square lattice with only
nearest neighbor couplings ($ J_2 =0 $).  This model is very interesting in
its own right, because the ground state of this model is probably
disordered having no long range order or it is at least very close
to a quantum phase transition,
in spite of the fact that the
lattice is bipartite and completely two-dimensional.
For this lattice structure, there are two topologically
inequivalent bonds.  One type of bonds which will be called as
plaquette-bonds hereafter form a plaquette covering of all
spins, Fig. 2. We introduce an exchange coupling, $J$ for this
type of bonds.
The other type of bonds, dimer-bonds, form a complete dimer covering
of the lattice and we use $J'$
for the exchange coupling of this type.
Although our final aim is to discuss the ground
state of the model with $J=J'$, for the time being we consider
$J$ and $J'$ as independent parameters. In comparison, all bonds
are equivalent in the square lattice. A dimer or plaquette covering
is possible in this case too. However, in contrast to the above
lattice it is not unique and it is known that the square lattice
favors an antiferromagnetically ordered
ground state.

We begin our discussion from the limit $J \gg J'$.
In this limit, as we have discussed (Table I with $J_1=J$ and $J_2=0$),
four spins on each plaquette form a singlet ground state which
has a character of resonating valence bond\cite{RVB}.
The ground-state energy of this
state per spin is
$  E_{\rm PRVB} = -\frac{1}{2}J$.
In the other limit, $J' \gg J$, the global ground-state is a
collection of dimer singlets,
$  E_{\rm dimer} = -\frac{3}{8}J'$.
For $J \approx J'$, another possible ground state is one with
antiferromagnetic long-range order, since the lattice is bipartite.
The energy of the classical N\'{e}el state is
$  E_{\rm N\acute{e}el} = - \frac{1}{8}J' - \frac{1}{4}J$.
These energies are plotted in Fig. 3.
The three states discussed here may be considered as the simplest
variational states for three different phases.  At this level,
the spin disordered phase, either the dimer phase or the PRVB phase,
has a lower energy than the N\'{e}el state.
Concerning the two singlet phases it should be mentioned that they
are different phases although the global symmetry properties are
the same for the two phases. It can be shown that
the wave function of the dimer singlet state has no overlap with
that of the PRVB state based on the fact that
the two wave functions have different transformation properties
under the reflection with respect to the dimer-bonds: odd for the
dimer singlet while even for the PRVB singlet. Thus, the two states
cannot be continuously connected by tuning $ J $ and $ J' $ from one
to the other limit.

It is necessary to improve the estimate of the ground-state
energies for the three phases.
For the singlet phases we can use perturbation theories.
Let us take the example of the PRVB state.  When a dimer-bond
is introduced, polarization processes from the singlets on both
ends of the bond should be included.
Polarization energy per bond may be calculated by the second
order perturbation as $-(43/576)(J'^2/J)$.  Similar perturbation
caluculation is also possible from the other limit of dimer singlet.
The results are summarized as
\begin{eqnarray}
  E_{\rm PRVB}  & = & -\frac{1}{2}J[1
                  +\frac{43}{576}(\frac{J'}{J})^2]\ ,\\
  E_{\rm dimer} & = & -\frac{3}{8}J'[1+\frac{1}{4}
                    (\frac{J}{J'})^2]\ .
\end{eqnarray}
These energy are also plotted in Fig.3.

For the N\'{e}el ordered phase a possible
improvement is obtained by
the linear spin wave theory\cite{anderson}. Extension of the linear
spin wave theory to the present case is rather complicated but straightforward.
The ground-state energy per spin in this approximation is
\begin{equation}
  E_{\rm N\acute{e}el}(J=J') = -J[\frac{3}{2}s^2+0.325248s]\ ,
\end{equation}
where $s=1/2$ for the present model.
At $J = J'$ the energy obtained by the spin wave theory, $-0.5376J$,
is very close to that of the PRVB state estimated
by the second order perturbation, $-0.5373J$.
In the spin wave theory it is possible to calculate
the reduction of magnetic moment by the zero point
fluctuations of the magnons, $\delta s =
0.288$ which is nearly 50 \% larger than the reduction for the
square lattice, $\delta s = 0.197$ and amounts to 58\% of the magnitude
of spin.
The linear spin wave theory shows that the N\'{e}el order survives
at this level but is on the verge of quantum phase transition.
In view of the fact that the linear spin wave theory has
a tendency to favor the N\'{e}el order\cite{hida}, more careful treatments
are necessary.

If there is a transition from the disordered phase to the N\'{e}el
phase it would be a second order transition.  In this case the spin
gap vanishes at the transition point.
Therefore the critical point
may be estimated by examining at the spin gap in the disordered phases.
A spin-triplet excitation in a plaquette is mobile. It can hop to
a neighboring plaquette with an effective hopping matrix element,
$J'/6$, to a second neighbor plaquette with $-J'^2/36J$ and
to a third neighbor plaquette with $-J'^2/216J$.
The polarization energies for the bonds connected to the
triplet are different form the polarization energy in the
ground state, $-(289/3456)(J'^2/J)$. From these results
the spin gap is calculated as
\begin{equation}
   \Delta_{\rm PRVB}  =  J[1-\frac{2}{3}(\frac{J'}{J})
                     -\frac{111}{864}(\frac{J'}{J})^2]\ .
\end{equation}
Similar second order perturbation gives the spin gap for the
dimer phase,
\begin{equation}
   \Delta_{\rm dimer}  =  J'[1-\frac{J}{J'}
                      -\frac{1}{2}(\frac{J}{J'})^2]\ .
\end{equation}
Within the second order perturbation theory, the spin gap vanishes at
$(J'/J)_c=1.215$ from the PRVB side and at $(J/J')_c=0.732$ from the
dimer side.  These points are shown by the dots in Fig. 3. This
result suggests that in the
narrow region between these critical points, antiferromagnetic
long range order would exist. However, the spin gap remains finite
at $J=J'$ within the present perturbation theory.

An alternative way to estimate the critical points is a cluster mean
field theory. We explain the method by a simple example.
For the PRVB state we consider a cluster with four spins on a
plaquette under the influence of molecular fields coming from
the dimer-bonds,
\begin{eqnarray}
   {\cal H}_{\rm CMF} =  J ({\bf s}_1+{\bf s}_3)\cdot
   ({\bf s}_2+{\bf s}_4)
    - J'\sigma(s_{1}^z -s_{2}^z+s_{3}^z-s_{4}^z)\ .
\end{eqnarray}
In the cluster mean field theory the average of a spin is determined by
the self-consistence equation, $\sigma=\langle s_{1}^z\rangle$.
This four spin problem can be solved analytically and
the critical value is obtained as $(J'/J)_c=3/4$.  When a
bigger cluster of 16-spins is used the critical value increases
to $(J'/J)_c=0.8044$.  One can use a similar cluster mean field
approximation for the dimer singlet.  The smallest cluster of two
spins gives $(J/J')_c=1/2$.  The next one of 8-spin cluster gives
$(J/J')_c=0.5378$.  From both sides the critical value increases
as the cluser size becomes larger.  Therefore we may consider
$(J'/J)_c=3/4$ and $(J/J')_c=1/2$ as the lower limits for
the critical points, if any.
However, unfortunately, cluster
size is not big enough to perform a reliable finite size scaling to
determine the existence or absence of the N\'{e}el phase.

All treatments discussed above suggest that the spin-1/2 Heisenberg model
on the 1/5-depleted lattice with only nearest neighbor couplings has
the spin disordered ground state in the wide region of parameter space.
Only in a narrow region around the crossing point between the PRVB phase
and the dimer phase, there is a possibility of the antiferromagnetic
long range order.
The recent quantum Monte Carlo simulations by Katoh and Imada \cite{katoh}
suggests that there
is a spin gap of $\Delta=0.11J$ for the model with $J =J'$
consistent with the perturbation results.

Let us return to the original model, Eq.(1), keeping the different
exchanges for the dimer-bonds and the plaquette bonds.  It should be
noted that the second nearest neighbor coupling is frustrating for
the N\'{e}el order.  It may be best illustrated by considering
the cluster mean field theory.
Again we consider the smallest cluster for the PRVB singlet,
\begin{eqnarray}
   {\cal H}_{\rm CMF}& = & J ({\bf s}_1+{\bf s}_3)\cdot
   ({\bf s}_2+{\bf s}_4) + J''({\bf s}_1\cdot{\bf s}_3
   +{\bf s}_2\cdot{\bf s}_4)
   \nonumber\\
   & & - (J'-2J'')\sigma(s_{1}^z -s_{2}^z+s_{3}^z-s_{4}^z)\ .
\end{eqnarray}
Since the eigenstates without the molecular field are completely
determined by quantum numbers listed in Table I, the critical value
is obtained analytically as
$(\frac{J' - 2J''}{J})_c=\frac{3}{4}$.
Thus we may conclude that the model for CaV$_4$O$_9$  has the quantum
disordered ground state with safe margin.

We can extend the second order perturbation theory for the spin
gap of the PRVB state with the frustating exchange coupling, for
which we obtain
\begin{eqnarray}
  \Delta_{\rm PRVB} &=& J \{1-\frac{2}{3}(x-2y)
    -\frac{1}{54}(7x^2-10xy+10y^2) \nonumber \\
  &&-\frac{1}{12}\frac{2x^2-3xy+4y^2}{2-y}
    +\frac{7}{18}\frac{(x-y)^2+y^2}{3-y} \\
  &&-\frac{1}{12}\frac{x^2+2y^2}{3-2y}
    -\frac{5}{72}\frac{(x-y)^2+y^2}{4-y}\}\ ,\nonumber
\end{eqnarray}
where $x=J'/J$ and $y=J''/J$.
 From this result it is seen that the gap increases as $J''$ increases.
This behavior is illustrated in Fig. 4 for the case $J=J'$.
A similar behavior is observed in the expansion around the dimer
limit which gives,
\begin{equation}
  \Delta_{\rm dimer} = J' \{ 1 - x^{-1}(1 - \frac{3}{2}y)
  - \frac{1}{8} x^{-2} (4 - 4y + 9y^2) \}.
\end{equation}
It is clear that quantitative results of the perturbation theory
is questionable at $J = J' = J_1 $,
which is seen by the difference between the
values of the spin gap for the model with only nearest neighbor
coupling.  $\Delta = 0.205J_1$ is obtained by the PRVB
perturbation theory on one hand and by the Quantum Monte
Carlo simulations, $\Delta = 0.11J_1$ on the other hand \cite{katoh}.
When we use the results of the Quantum Monte Carlo simulations
the magnitude of the spin gap is too small compared with the
experimental value.  The present perturbation result
shows that the spin gap increases significantly
when we include the frustrating next nearest neighbor exchange
of the order of
$J_2=J_1/2$, which may lead to a reasonable value of the spin gap
compared with the experiments.

In conclusion the spin-1/2 Heisenberg model on the 1/5-depleted
square lattice is presented as a theoretical model for the spin
gap of CaV$_4$O$_9$.  It is shown that the 1/5 depleted square lattice
is favorable for the quantum spin disordered phase, which may
be characterized as the PRVB singlet.
When the frustrating exchange for the corner sharing bonds is
included it is possible to explain the large spin gap observed
experimentally.

\acknowledgments
It is our great pleasure to acknowledge fruitful discussions
with Masatoshi Sato, Hirokazu Tsunetsugu, Yoshio Kitaoka, and
Maurice Rice.  This work is financilly supported by a Grant-in-Aid
>from the Ministry of Education, Science and Culture of Japan.
We acknowledge financial support from Swiss Nationalfonds (M.S.;
(No. 8220-037229) and from M.I.T. Science Partnership Fund and from
the NSF (P.L. and M.S.; DMR-90-22933).

A significant part of this work was done during the stay in
Aspen Center for Physics of K.U. and M.S.

\begin{figure}
\caption
{A model for CaV$_4$O$_9$ of the spin-1/2 Heisenberg model
with the nearest neighbor (solid lines) and
the next nearest neighbor (broken lines) exchange interactions.
The dot dashed lines show the unit cell of the 1/5 depleted
square lattice.}
\label{model}
\end{figure}

\begin{figure}
\caption
{Spin 1/2 Heisenberg model on the 1/5-depleted square lattice
with the nearest neighbor exchange interactions.  Topologically
there are two different type of bonds: plaquette-bonds $J$
and dimer-bonds $J'$.}
\label{DSL}
\end{figure}

\begin{figure}
\caption
{Ground state energies for the dimer singlet, the PRVB siglet, and
the N\'{e}el ordered phase.  The horizontal axis is $ \tilde{x} $ defined by
$J/J'= \tilde{x}/(1- \tilde{x})$. The points where the spin gap vanishes
are denoted by the dots. $E_{SW}$ is the energy estimated by the linear spin
wave theory at $J=J'$}
\label{energy}
\end{figure}

\begin{figure}
\caption
{The spin gap as a function of frustrating exchange coupling,
$y=J''/J$. For $J''/J > 0.354$ the spin gap deviates from Eq.(10)
because the minimum of the spectrum is different from $(\pi,\pi)$.}
\label{gap}
\end{figure}


\begin{references}
\bibitem{dxy} Another possibility is the $d_{xy}$ orbital
since Vanadium ion is shifted from the basal plane towards inside
of the pyramid of oxgens.  Even if this is the case the following
discussions about the superexchange remains qualitatively the same
with the use of $p_x$ and $p_y$ orbitals.
\end{references}
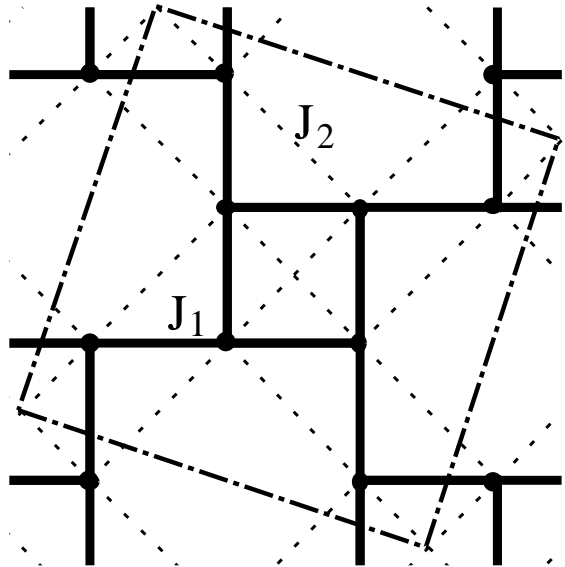
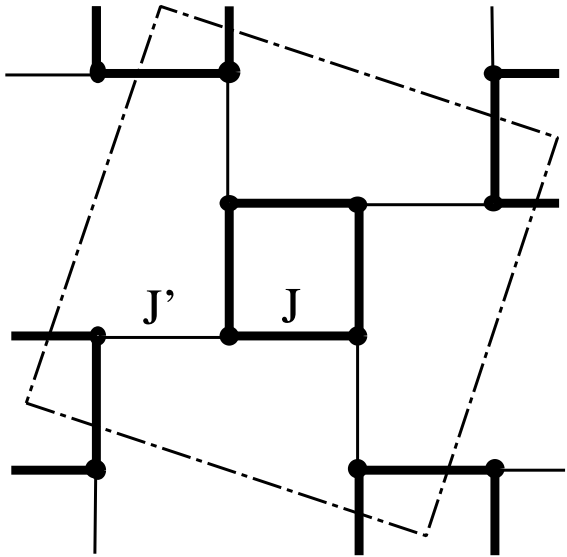
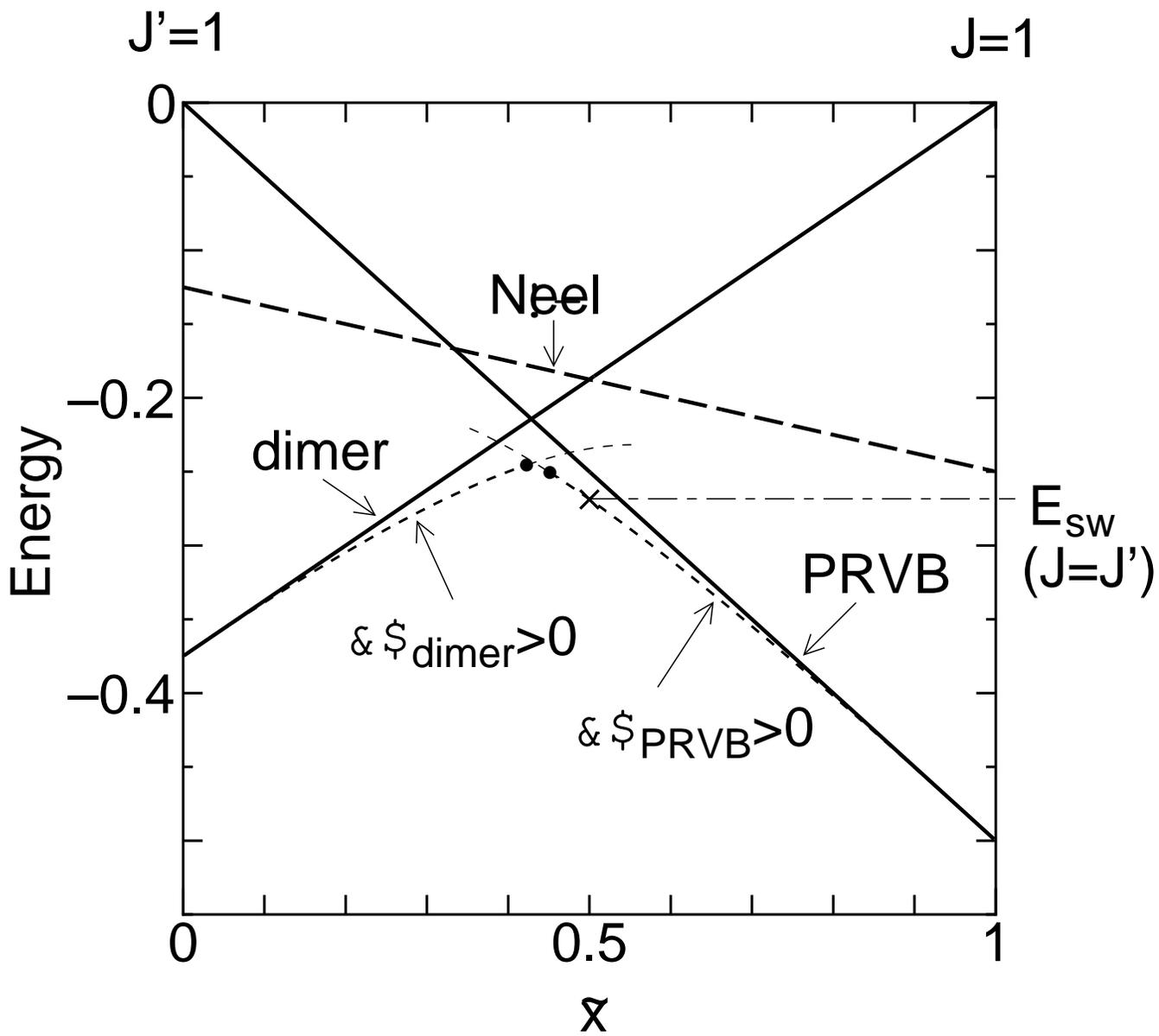
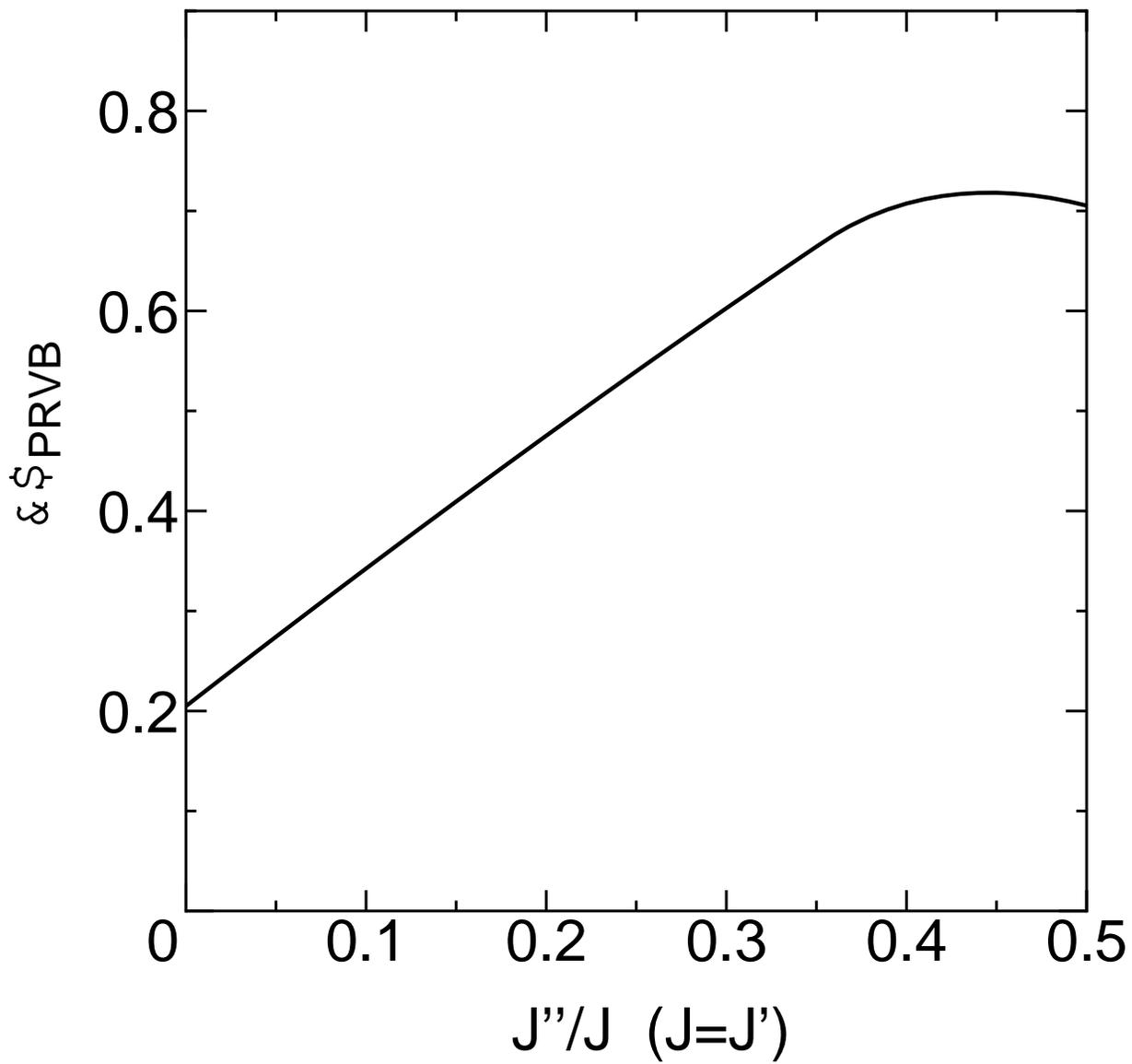
\end{document}